\documentclass[preprint,prc,aps,nofootinbib,superscriptaddress]{revtex4}
\usepackage{graphicx}
\usepackage{epsfig}
\usepackage{bm}
\usepackage{fancybox}
\usepackage{amssymb}

\newcommand{\lag}{\mathcal L}

\begin{document}

%\title{Kaon optical potential in nuclei and kaon condensation in neutron star}
\title{Kaon condensation and composition of neutron star matter in modified quark-meson coupling model}

\author{C. Y. Ryu} \email{cyryu@rcnp.osaka-u.ac.jp}
\affiliation{Department of Physics and Basic Atomic Energy Research Institute, 
Sungkyunkwan University,
Suwon 440-746, Korea}

\author{C. H. Hyun} \email{hch@meson.skku.ac.kr}
\affiliation{Department of Physics and Institute of Basic Science, 
Sungkyunkwan University, Suwon 440-746, Korea}
\affiliation{School of Physics, Seoul National University,
Seoul 151-742, Korea}

\author{S. W. Hong} \email{swhong@skku.ac.kr}
\affiliation{Department of Physics and Institute of Basic Science, 
Sungkyunkwan University, Suwon 440-746, Korea}

\author{B. T. Kim} \email{btkim@skku.ac.kr}
\affiliation{Department of Physics and Institute of Basic Science, 
Sungkyunkwan University, Suwon 440-746, Korea}

\date{March 9, 2007}

\begin{abstract}

We use the modified quark-meson coupling (MQMC) model to study
the composition profile of neutron star matter and compare the results
with those calculated by quantum hadrodynamics (QHD).
Both MQMC and QHD model parameters are adjusted to produce exactly
the same saturation properties so that we can investigate the model
dependences of the matter composition at high densities.
We consider the possibility 
of deep kaon optical potential
and find that the composition of matter is very sensitive to
the interaction strength of kaons with matter. 
The onset densities of the kaon condensation are studied in detail
by varying the kaon optical potentials.
We find that the MQMC model produces the kaon condensation 
at lower densities than QHD.
The presence of kaon condensation
changes drastically the population of octet baryons and leptons.
Once the kaon condensation takes place, the population of
kaons builds up very quickly, and kaons become the dominant component
of the matter.
We find that the $\omega$-meson plays an important role in
increasing the kaon population and suppressing the hyperon population.
\end{abstract}

\maketitle

\section{Introduction}

Observation of neutron star properties such as 
mass, size and temperature
provides us with important clues to the understanding of the state of matter
at extremely high densities. 
In the 1970's, the maximum mass of the neutron star
was calculated with the $NN$ potentials available at that time 
\cite{pand-npa71,bj-npa74,ps-npa75,fp-npa81}
and mean field models \cite{wal-ap74,ps-plb75}. 
Most of the calculations done in the 1970's resulted in 
stiff equations of state, 
and thus the maximum mass of 
a neutron star was predicted to be larger than $2 M_\odot$,
where $M_\odot$ is the solar mass.
Only the Reid soft core potential yielded a soft equation of 
state and consequently a small maximum mass of a neutron star,
$1.6 M_\odot$ \cite{pand-npa71}.
Recent observations of the masses of binary pulsars \cite{tc-apj99},
which are candidates of neutron stars indicate that the
maximum mass of neutron stars are roughly around $1.5 M_\odot$,
substantially smaller than most of the predicted values in the 1970's. 
(However, very recent observations seem to suggest possible
existence of more massive pulsars in the range $(1.8 - 2.0) M_\odot$
\cite{apj2004,apj2005}, though further confirmation is needed.)
On the other hand, exotic forms of matter, i.e., matter consisting
of degrees of freedom other than the nucleons have been proposed 
already many years ago.
Some of the proposed exotic states of matter include those with
creation of hyperons \cite{hyperon}, 
Bose-Einstein condensation (pions \cite{pion} or kaons \cite{kaon}), 
strange matter \cite{strange}, and 
quark deconfinement \cite{itoh70,bc-plb76,kk-plb76}.
These exotic states seem to reduce the maximum mass of a neutron star 
close to the observations 
\cite{kpe-prc95, s-pal, nkg-prd92}, 
implying that exotic degrees of freedom seem to be needed
to reproduce the observed masses of neutron stars.

In this work, we consider the strangeness degrees of freedom by including
both hyperon creation and kaon condensation in the neutron star matter.
(It is the anti-kaon that matters here, but we simply refer to
both kaons and anti-kaons as kaons for brevity.)
The masses and energies of the hyperons and kaons in medium are sensitive to
their interactions with the surrounding matter. 
In the meson-exchange picture,
meson-hyperon and meson-kaon coupling constants can fix the
strength of these interactions. 
The meson-hyperon coupling constants may be
determined from the binding energies of hyperons in hypernuclei.
The meson-kaon coupling constants have been studied by using
the kaon-nucleon scattering \cite{Cieply,Kaiser} 
and kaonic atom data \cite{Cieply}.
Recently, the magnitudes of the kaon-nucleus potential 
in matter have attracted much attention.
Some calculations \cite{Schaffner,oset20,Cieply}
show that the real part of the $K^-$-nucleus optical potential $U_{K^-}$ 
is shallow ($U_{K^-} \approx$ $-$50 MeV), but some other calculations
suggest that $U_{K^-}$ can be as large as about $-120$ MeV 
\cite{gal94,Kaiser} or even close to $-200$ MeV \cite{Batty}.

Akaishi and Yamazaki predicted possible existence of
deeply bound kaonic nuclei \cite{akaishi},
in which $U_{K^-}$ at normal density $\rho_0$ was
estimated to be about $-120$ MeV.
Then, experiments at KEK claimed the observation of 
tribaryon kaonic nuclei, 
S$^0$ \cite{s0} and S$^+$ \cite{suzuki1}, which seem to suggest that 
$K^-$ may be even more deeply bound than 
the theoretical prediction \cite{akaishi}
(The former claim \cite{s0}, however, was withdrawn by the experimental group
\cite{iwasaki}). 
FINUDA collaboration at DA$\Phi$NE \cite{FINUDA} 
and a BNL experiment with $^{16}{\rm O}(K^-, n)$ reaction \cite{kishi-npa05} 
also reported distinct peaks. 
More recently there was a theoretical work which considered
large kaonic binding energies and calculated widths of kaonic nuclear
bound states \cite{Mares}.
The identities of these experimental peaks need to be
studied further experimentally and theoretically. However, in this work, 
we consider the possibility of  
deep optical potential of kaons in nuclei and 
explore the consequences in the 
composition profile of neutron star matter.

In this work, for the description of dense matter
we employ the modified quark-meson coupling (MQMC) model 
\cite{mqmc}.
Nucleons and hyperons in the baryon octet are treated as MIT bags.
The bag constant $B_B$ and phenomenological constant $Z_B$ for a baryon 
$B$ are fixed to reproduce the free mass of each baryon $B$.
Coupling constants between ($u$, $d$)-quarks in the bags and ($\sigma$, 
$\omega$, $\rho$)-mesons are adjusted to give us the 
binding energy per a nucleon $E_b/A = 16$ MeV and symmetry energy  
$a_{\rm sym} = 32.5$ MeV at the
saturation density $\rho_0 = 0.17\, {\rm fm}^{-3}$. 
Since the interaction between the $s$-quark and mesons are not well known,
we adopt the standard quark counting rule and assume the $s$-quark is
decoupled from ($\sigma$, $\omega$, $\rho$)-mesons.
To take into account the interactions between $s$-quarks,
we introduce $\sigma^* (980)$ and $\phi(1020)$ mesons
following Ref.~\cite{s-pal} for the baryon and Ref.~\cite{bb-prc01} 
for the kaon.
We also assume the kaon as a point particle.
This treatment allows us to use $U_{K^-}$
as an input to fix the coupling constant between the $\sigma$-meson 
and the kaon, $g_{\sigma K}$. 
In our model the real part of the kaon optical potential at $\rho = \rho_0$
can be written as
$U_{K^-} = - (g_{\sigma K} \sigma(\rho_0)
+ g_{\omega K} \omega(\rho_0)) $,
where $\sigma(\rho_0)$ and $\omega(\rho_0)$ are the values of
the meson fields at $\rho_0$. Using the value of $g_{\omega K}$ 
given by the quark counting rule,
we can fix $g_{\sigma K}$ for each given value of $U_{K^-}$. 
Once the parameters of the model are fixed,
the composition profile of neutron star matter can be obtained from
the $\beta$-equilibrium and charge neutrality.
We find that the composition of neutron star matter changes 
dramatically depending on the value of $U_{K^-}$.

To investigate the model dependence of the results
we also employ quantum hadrodynamics (QHD)
model \cite{sw-qhd} for calculating the composition of matter.
The parameters of the QHD model are calibrated to produce exactly 
the same saturation properties as in the MQMC model.
Our calculations show that the onset densities of 
the kaon condensation and the compositions of matter
at high densities are substantially model dependent.
In Sect. II, we introduce model Lagrangians and fix the model parameters.
The results are discussed in Sect. III.  
Conclusions and discussions follow in Sect. IV.

\section{Theory}

In this section we first briefly sketch the MQMC and QHD models
by presenting the model Lagrangians.
The models are calibrated so as to be consistent with each other
at the saturation density by fixing the 
coupling constants of both models to produce exactly
the same saturation properties; the saturation density,
the binding energy, the symmetry energy, the nucleon effective
mass, and the compression modulus.
We then show how the physical quantities that will determine 
the composition of the neutron star matter can be obtained self-consistently.
\subsection{{\bf Models}}
\label{sect:The MQMC model with kaons}

The model Lagrangian comprises
the terms for the octet baryons, exchange mesons, leptons and kaons;
$\lag_{tot} = \lag_B + \lag_M + \lag_l + \lag_K$.
Octet baryon, exchange meson and lepton terms 
in the mean field approximation can be written as
\begin{eqnarray}
\lag_B  & =& \sum_B \bar \psi_B \left[i\gamma \cdot \partial - m_B^* 
(\sigma, \sigma^*) %\nonumber\\ && 
-\gamma^0 \left(g_{\omega B}\omega_0 + g_{\phi B}\phi_0 + \frac 12 g_{\rho B}
\tau_z \rho_{03} \right) \right] \psi_B \label{eq:lagb} \\ 
\lag_M &=& -\frac 12 m_\sigma^2 \sigma^2 - \frac 12 m_{\sigma^*}^2 {\sigma^*}^2
+ \frac 12 m_\omega^2 \omega_0^2 + \frac 12 m_\phi^2 \phi_0^2 + \frac 12 m_\rho^2 \rho_{03}^2, \label{eq:lagm} \\ 
\lag_l &=& \sum_l \bar \psi_l ( i \gamma \cdot \partial - m_l)\psi_l ,
\label{eq:lagl}
\end{eqnarray}
where $B$ denotes the sum over all the octet baryons 
($p, ~n, ~\Lambda, ~\Sigma^+, ~\Sigma^0, ~\Sigma^-, ~\Xi^0, ~\Xi^-$),
%coupled to the ($\sigma$, $\omega$, $\rho$, $\sigma^*$, $\phi$) mesons
and $l$ stands for the sum over the free electrons and muons ($e^-$, $\mu^-$).
$\sigma$, $\omega$ and $\rho$ mesons mediate the interactions
between the non-strange light quarks ($u$ and $d$).
$\sigma^*$ and $\phi$ mesons are introduced to take into
account the interactions between $s$ quarks.
$\lag_B$ is of the identical form for both 
the MQMC and QHD models,
but differs in the definition of
the effective baryon mass $m^*_B$
as will be shown below. 

{\bf MQMC}

In the MQMC model, a baryon is a composite system with quarks 
in a spherical bag, and its mass is given in terms of bag
parameters and quark eigenenergy.
The effective mass of a baryon in matter $m^{*}_B(\sigma, \sigma^*)$
can be written as \cite{s-pal,mqmc,fleck,st,stt}
\begin{eqnarray}
m^*_B = \sqrt{E^2_B - \sum_q  \left(\frac{x_q}{R} \right)^2}.
\label{eq:efmass}
\end{eqnarray}
The bag energy of a baryon is given by 
\begin{eqnarray}
E_B &=& \sum_q  \frac{\Omega_q}{R} - \frac{Z_B}{R}
+ \frac{4}{3} \pi\, R^3\, B_B,
\label{eq:bagery}
\end{eqnarray}
where $B_B$ and $Z_B$ are the bag constant and  
a phenomenological constant for the zero-point motion of a baryon $B$,
respectively.
$\Omega_q = \sqrt{x^2_q + (R m^*_q)^2}$, 
where $m^*_q (= m_q - g^q_\sigma \sigma - g^q_{\sigma^*} \sigma^*)$
is the effective mass of a quark whose free mass is $m_q$.
We take $m_q =0$ for $q=u,d$ and $m_q = 150$ MeV for $q=s$.
(Other choices of $m_{q=s}$ values do not make differences in the 
results~\cite{Theta+}.)
$x_q$ is determined from the boundary condition on the bag surface $r = R$,
\begin{equation}
j_0(x_q) = \beta_q j_1(x_q),
\label{eq:bessel}
\end{equation}
where %$j_0 (x_q)$ and $j_1 (x_q)$ are the spherical bessel functions and
$
\beta_q = \sqrt{\frac{\Omega_q - R m^*_q}{\Omega_q + R m^*_q}}.
$
In the MQMC model, the bag constant $B_B$ is assumed to depend on density
\cite{mqmc,stt}.
In this work, we use the extended form in Ref.~\cite{s-pal} to include the 
contribution from $\sigma^*$ as
\begin{eqnarray}
B_B (\sigma,\, \sigma^*)  =  
B_{B0} \exp \left\{ -\frac{4}{m_B} \left({g'}_\sigma^B \sum_{q=u,d} n_q \sigma
+ {g'}_{\sigma^*}^B (3-\sum_{q=u,d}n_q) \sigma^*\right) \right\},
\label{eq:bag}
\end{eqnarray}
where $m_B$ is the free mass of the baryon $B$,
%As noted in the Introduction, 
assuming that the $\sigma$ meson couples to $u$ and $d$ quarks only
and that the $\sigma^*$ meson couples to the $s$ quark only.
%The factor $\sqrt{2}$ in front of $\sigma^*$ is introduced due to
%SU(6) symmetry \cite{s-pal}.

{\bf QHD}

In the QHD model, a baryon is treated as a point particle,
and thus its effective mass is simply written as 
\begin{eqnarray}
m^*_B = m_B - g_{\sigma B} \sigma - g_{\sigma^* B} \sigma^*.
\end{eqnarray}
In order to reproduce the same saturation properties 
as obtained in the MQMC model, 
self-interactions of the $\sigma$-field \cite{bb-npa77}
\begin{eqnarray}
U^{\rm QHD}_{\sigma} = \frac{1}{3} g_2\, \sigma^3 + 
\frac{1}{4} g_3\, \sigma^4 
\label{eq:U_QHD}
\end{eqnarray}
are added to Eq.~(\ref{eq:lagm}) so that
\begin{eqnarray}
\lag_M^{\rm QHD} = \lag_M - U^{\rm QHD}_{\sigma}.
\end{eqnarray}
As mentioned above, the baryon and the lepton Lagrangians for the QHD model 
take the form given by 
Eqs.~(\ref{eq:lagb}) and (\ref{eq:lagl}).

{\bf Kaon}

The effective Lagrangian for the kaon may be expressed as \cite{glend}
\begin{eqnarray}
\lag_K = D_\mu^* K^* D^\mu K - {m_K^*}^2 K^* K,
\label{eq:lagk}
\end{eqnarray}
where 
$
D_\mu = \partial_\mu + i g_{\omega K}\omega_\mu 
-i g_{\phi K} \phi_\mu + i \frac 12 g_{\rho K} \vec \tau \cdot \vec \rho_\mu.
$
In this work we treat the kaon
as a point particle in both MQMC and QHD models,
and its effective mass is given by
\begin{equation}
m_K^* = m_K - g_{\sigma K} \sigma - g_{\sigma^* K} \sigma^*.
\label{eq:keffmass}
\end{equation}
The equation of motion for a kaon is given by   
\begin{eqnarray}
[D_\mu D^\mu + {m_K^*}^2] K(x) = 0.
\end{eqnarray}
In uniform infinite matter, the kaon field $K(x)$ 
can be written as a plane wave.
Substituting the plane wave solution into the equation 
of motion, we obtain the dispersion relation for the anti-kaon
\begin{eqnarray}
\omega_K = m_K^* - g_{\omega K} \omega_0 
+ g_{\phi K} \phi_0 - g_{\rho K} \frac 12 \rho_{03}.
\label{eq:kaon-dispersion}
\end{eqnarray}
%where $I_{3K}$ is the isospin third component of the anti-kaon.

\subsection{Model parameters}

{\bf MQMC}

In the MQMC model, MIT bag parameters $B_{B0}$ and $Z_B$ are determined
to reproduce the free mass of a baryon $B$,
$m^*_B \left|_{\rho = 0} = m_B \right. $ 
with the minimization condition 
$\left.  \frac{\partial m_B}{\partial R} \right|_{R = R_0} =0$
at a free bag radius $R_0$, which we choose as $R_0 = 0.6$ fm.
The bag parameters $B_{B0}$ and $Z_B$ for the octet baryons
are listed in Table~\ref{tab:b-z}.
\begin{table}[tbp]
\begin{center}
\begin{tabular}{|c|c|c|c|} \hline
~~ $B$ ~~  &  $m_B$ (MeV) & $B^{1/4}_{B0}$ (MeV) & $Z_B$ \\ \hline
   $N$     &   939.0 & ~~188.1 ~~& ~~2.030 ~~\\ \hline
$\Lambda$  &  1115.6 &   197.6   &   1.926   \\ \hline
$\Sigma^+$ &  1189.4 &   202.7   &   1.829   \\ \hline
$\Sigma^0$ &  1192.0 &   202.9   &   1.826   \\ \hline
$\Sigma^-$ &  1197.3 &   203.3   &   1.819   \\ \hline
$\Xi^0   $ &  1314.7 &   207.6   &   1.775   \\ \hline
$\Xi^-   $ &  1321.3 &   208.0   &   1.765   \\ \hline
\end{tabular}
\end{center}
\caption{Bag constants $B_{B0}$ and phenomenological constants $Z_B$ for
octet baryons to reproduce the free mass of each baryon.  
The bag radius is chosen as $R_0=0.6$ fm for all octet baryons, 
and the bare masses of quarks are fixed as
$m_{u(d)}=0$ MeV and $m_s = 150$ MeV.}
\label{tab:b-z}
\end{table}

Three saturation
conditions $\rho_0$, $E_b/A$, and $a_{\rm sym}$ could determine
three quark-meson coupling constants $g^{u,d}_\sigma$,
$g^{u,d}_\omega$ and $g^{u,d}_\rho$, assuming $u$ and $d$ quarks
to be identical in the isodoublet.
The MQMC model, however, introduces an additional constant
$g'^B_\sigma$ in Eq.~(\ref{eq:bag}). 
Thus we fix $g^{u,d}_\sigma = 1$, and adjust the remaining
three constants to meet the three conditions.
The resulting coupling constants are given in Table~\ref{tab:coupling}
together with the ratio of the effective mass of the nucleon $m^*_N/m_N$
and the compression modulus $K$.
\begin{table}
\begin{center}
\begin{tabular}{|c|c|c|c||c|c|} \hline
~~$g_\sigma^q$~~  & ~~$g_\omega^q$~~   & ~~${g'}_\sigma^B$~~  & 
~~$g_\rho^q$~~    & $m_N^* / m_N$  & $K$ (MeV)  \\ \hline
         1.0      & 2.71           & 2.27       &
         7.88     & 0.78           & 285.5      \\ \hline
\end{tabular}
\end{center} 
\caption{The coupling constants between $(u,\, d)$-quarks and 
$(\sigma,\, \omega,\, \rho)$-mesons in the MQMC 
model to reproduce the binding energy $E_b/A=16$ MeV 
and symmetry energy $a_{\rm sym}=32.5$ MeV
at the saturation density 0.17 ${\rm fm}^{-3}$. 
$m^*_N/m_N$ and $K$ are the ratio of the 
effective mass to the free mass of 
the nucleon and the compression modulus at the saturation density,
respectively.}
\label{tab:coupling}
\end{table}
$m^*_N$ and $K$ are within reasonable ranges: 
$m^*_N = (0.7 \sim 0.8) m_N$ and $ K = (200 \sim 300)$ MeV.

The coupling constants between $s$-quarks and mesons
cannot be determined from the
saturation properties.
In principle, experimental data from hypernuclei 
and kaon-nucleus scattering could be used to determine 
the coupling constants between $s$-quarks and mesons (for example,
see Ref.~\cite{zakout05}).
However, these coupling constants are not well known yet, 
and for simplicity
we assume that the quark counting rule holds and that
the $s$ quark does not interact with $u$ and $d$ quarks.
Then we have
\begin{equation}
g^s_{\sigma} = g^s_{\omega} = g^s_{\rho} = 
g^{u,d}_{\sigma*} = g^{u,d}_\phi = 0.
\label{eq:smcoupling}
\end{equation}
To fix the meson-baryon coupling constants in the model Lagrangian,
we also use the quark counting rule 
\begin{eqnarray}
\frac 13 g_{\omega N} & = & \frac 12 g_{\omega \Lambda}
= \frac 12 g_{\omega \Sigma} =g_{\omega \Xi} = g_\omega ^q , \nonumber \\
g_{\rho N} & = & g_{\rho \Sigma} = g_{\rho \Xi} = g_\rho^q , 
~~g_{\rho \Lambda}=0, \nonumber \\
g_{\phi \Lambda} & = & g_{\phi \Sigma} = \frac 12 g_{\phi \Xi} = g_\phi^s,
\label{eq:211}
\end{eqnarray}
and the SU(6) symmetry
\begin{eqnarray}
g^s_{\sigma*} &=& \sqrt{2} g^{u,d}_\sigma = \sqrt{2}, \nonumber \\
g_\phi^s &=& \sqrt 2 g_\omega^{u,d} = 3.83, \nonumber \\
{g'}^B_{\sigma^*} &=& \sqrt{2}\, {g'}^B_\sigma.
\label{eq:su6}
\end{eqnarray}
The quark-meson coupling constants $g^q _{\omega}$ and $g^q _{\rho}$
given in Table~\ref{tab:coupling} and
the relations in Eqs.~(15)-(17)
%the quark counting relations 
%in Eqs.~(\ref{eq:smcoupling}) and (\ref{eq:211}) 
determine all the meson-baryon coupling of the MQMC model.

{\bf QHD}

In the QHD model, $g_{\sigma N}$ and $g_{\omega N}$ are adjusted
to yield $\rho_0$ and $E_b$, and $g_{\rho N}$ 
is fitted to produce $a_{\rm sym}$.
$g_2$ and $g_3$ in $U^{\rm QHD}_{\sigma}$ of Eq.~(\ref{eq:U_QHD})
are fixed to 
reproduce the same $m^*_N$ and $K$ values as listed in 
Table~\ref{tab:coupling} for the MQMC model.
The coupling constants determined in this way
are given in Table~\ref{tab:coupling-qhd}.
\begin{table}[tbp]
\begin{center}
\begin{tabular}{|c|c|c|c|c|}\hline
~~$g_{\sigma N}$~~ & ~~$g_{\omega N}$~~ & ~~$g_{\rho N}$~~ &
$g_2$ (fm$^{-1}$) & $g_3$ \\ \hline
8.06 & 8.19 & 7.88 & 12.139 & 48.414 \\ \hline 
\end{tabular}
\end{center}
\caption{The meson-nucleon coupling constants and the coefficients
of the $\sigma$-meson self interaction terms used in the QHD model.
They reproduce the same saturation properties as in the MQMC model; 
$\rho_0 = 0.17$ fm$^{-3}$, $E_b = 16 A$ MeV, $a_{\rm sym} = 32.5$ MeV, 
$m^*_N = 0.78 m_N$ and $K = 285.5$ MeV.} 
\label{tab:coupling-qhd}
\end{table}
In the MQMC model, meson-baryon coupling constants are
obtained from the quark-meson coupling constants.
On the other hand, in QHD
meson-nucleon coupling constants provide the starting point
for the determination of remaining other meson-baryon
coupling constants.
Once meson-nucleon coupling constants are fixed from the saturation
properties, meson-hyperon coupling constants can be obtained 
by the quark counting rule (as in Eq.~(\ref{eq:211}))
and the SU(6) symmetry (as in Eq.~(\ref{eq:su6})).
The coupling constants between strange mesons and hyperons can be obtained
by combining the quark counting rule and the SU(6) symmetry,
{\it e.g.}, $g_{\phi \Lambda} = \sqrt{2} g_{\omega N} /3$ and
$g_{\sigma^* \Lambda} = \sqrt{2}g_{\sigma N}/3$.

{\bf Kaon}

There are 5 kaon-meson coupling constants in our models; $g_{\sigma K}$,
$g_{\omega K}$, $g_{\rho K}$, $g_{\sigma^* K}$ and $g_{\phi K}$.
$g_{\omega K}$ and $g_{\rho K}$ can be fixed
from the quark counting rule: $g_{\omega K} = g_\omega^q$ and 
$g_{\rho K}=g_\rho^q$
for the MQMC model, and $g_{\omega K} = g_{\omega N}/3$ and
$g_{\rho K}=g_{\rho N}$ for QHD. 
(Obviously $g_{\rho K}$ from the MQMC model is the same
as that from QHD. $g_{\omega K} (= 2.71)$ from the MQMC model is
essentially the same as $g_{\omega K} (=2.73)$ from QHD.)
$g_{\sigma^* K}$ may be fixed from $f_0(980)$ decay \cite{wa76-91}, 
and $g_{\phi K}$ from the SU(6) relation 
$\sqrt{2} g_{\phi K} = g_{\pi \pi \rho} = 6.04$ \cite{j-schaf}.  
$g_{\sigma^* K}$ and $g_{\phi K}$ thus fixed are 2.65 and 4.27,
respectively.
The remaining coupling constant, $g_{\sigma K}$, 
can be related to the real part of the optical potential 
of a kaon at the saturation density through
$U_{K^-} = -(g_{\sigma K}\sigma + g_{\omega K}\omega_0)$.
$g_{\sigma K}$ values corresponding to
several values of $U_{K^-}$ are listed in Table~\ref{tab:gsigmaK}
for both MQMC and QHD models.

Thus, out of 5 kaon-meson coupling constants
$g_{\rho K}$, $g_{\sigma^* K}$, and $g_{\phi K}$ are the same for  
both models. $g_{\omega K}$'s are essentially the same for both models.
%; 2.71 and 2.73 for the MQMC and QHD models, respectively.
$g_{\sigma K}$ values are also very similar 
in both models for all $U_{K^-}$ values as seen in Table IV.  
Therefore, all the 5 kaon-meson coupling constants are practically
the same for both MQMC and QHD models.
\begin{table}
\begin{center}
\begin{tabular}{|c|c|c|c|c|c|} \hline
$   U_{K^-}  $ (MeV)  & $-80$ & $-100$ & $-120$ & $-140$ & $-160$ \\ \hline
$g_{\sigma K}$ (MQMC) & 1.25  & 2.01   & 2.75   & 3.50   & 4.25   \\ \hline
$g_{\sigma K}$ (QHD)  & 1.26  & 2.04   & 2.82   & 3.61   & 4.39   \\ \hline
\end{tabular}
\end{center}
\caption{$g_{\sigma K}$ determined for several $U_{K^-}$ values
in the MQMC and QHD models.}
\label{tab:gsigmaK}
\end{table}

\subsection{Other quantities relevant to neutron star matter}  

To obtain the composition of neutron star matter, 
we need to determine 16 unknown variables at 
each matter density, which include 5 meson fields 
($\sigma , \omega , \rho , \sigma^* , \phi$ ), 
8 octet baryon densities, 2 lepton densities 
and the kaon density $\rho_K$.
Five meson fields can be determined from their equations of motion:
\begin{eqnarray}
m_\sigma^2 \sigma + \frac{\partial}{\partial \sigma} U^{\rm QHD}_{\sigma}
= \sum_B g_{\sigma B} C_B(\sigma)
\frac{2J_B+1}{2\pi^2} \int_0^{k_B} \frac{m_B^*}{[k^2+{m_B^*}^2]^{1/2}}k^2 dk
+ g_{\sigma K} \rho_K,
\label{sigma}
\end{eqnarray}
\begin{eqnarray}
m_{\sigma^*}^2 \sigma^* = \sum_B g_{\sigma^* B} C_B(\sigma^*)
\frac{2J_B+1}{2\pi^2} \int_0^{k_B} \frac{m_B^*}{[k^2+{m_B^*}^2]^{1/2}}k^2 dk
+ g_{\sigma^* K}\rho_K,
\label{sigmastar}
\end{eqnarray}
\begin{eqnarray}
m_\omega^2 \omega_0 = \sum_B g_{\omega B} (2J_B + 1) k_B^3 / (6\pi^2)
- g_{\omega K}\rho_K,
\label{eq:omega}
\end{eqnarray}
\begin{eqnarray}
m_\phi^2 \phi_0 = \sum_B g_{\phi B} (2J_B + 1) k_B^3 / (6\pi^2)
+ g_{\phi K}\rho_K,
\label{eq:phi}
\end{eqnarray}
\begin{eqnarray}
m_\rho^2 \rho_{03} = \sum_B g_{\rho B} I_{3B} (2J_B + 1) k_B^3 / (6\pi^2)
-g_{\rho K} \frac 12 \rho_K ,
\label{eq:rho}
\end{eqnarray}
where $J_B$ and $I_{3B}$ are the spin and the isospin projection, respectively, 
and $k_B$ is the Fermi momentum of the baryon $B$. 
In Eq.~(\ref{sigma}) $\frac{\partial} {\partial \sigma} U^{\rm QHD}_{\sigma}$ 
term needs to be there only for QHD 
and is not to be included in the MQMC model.
$C_B(\sigma)$ and $C_B(\sigma^*)$ are determined from the relations
$g_{\sigma B} C_B(\sigma) = - \frac{\partial m_B^*}{\partial \sigma}$
and
$g_{\sigma^* B} C_B(\sigma^*) = - \frac{\partial m_B^*}{\partial \sigma^*}$ .
For QHD, $C_B(\sigma) = C_B(\sigma^*) = 1$.  
For MQMC, the explicit forms of $C_B(\sigma)$ and $C_B(\sigma^*)$ 
are given in Ref. \cite{s-pal}.

Charge neutrality condition of neutron star matter is expressed as 
\begin{eqnarray}
\sum_B q_B \rho_B - \rho_K - \rho_e - \rho_\mu = 0,
\label{eq:baryonconserv}
\end{eqnarray}
where $q_B$ is the charge of baryon $B$ and $\rho_B$ is the number
density of $B$.
Using the charge neutrality and the baryon number conservation conditions,
one can fix two quantities, {\it e.g.,} the density
of the neutron and the electron.
With these two variables fixed,
$\beta$-equilibrium conditions of the baryons give us the following 7 relations
for the chemical potentials of 
$p,\, \Lambda,\, \Sigma^+,\, \Sigma^-,\, \Sigma^0,\, \Xi^-$
and $\Xi^0$ as
\begin{eqnarray}
\mu_n = \mu_\Lambda &=& \mu_{\Sigma^0} = \mu_{\Xi^0} , \nonumber \\
\mu_n + \mu_e &=& \mu_{\Sigma^-} = \mu_{\Xi^-} , \nonumber \\
\mu_n - \mu_e &=& \mu_p = \mu_{\Sigma^+},
\label{eq:chemi-eq}
\end{eqnarray}
where the chemical potential of baryon $B$ is given by
$
\mu_B = \sqrt{k_B^2 + {m_B^*}^2(\sigma,\sigma^*)} + g_{\omega B}\omega_0
+ g_{\phi B} \phi_0 + g_{\rho B} I_{3B}\rho_{03}.
$
The chemical potential of a non-interacting lepton $l$ is simply written as
$
\mu_l = \sqrt{k^2_l + m^2_l}.
$
The $\beta$-equilibrium condition for leptons
\begin{equation}
\mu_e = \mu_\mu 
\end{equation}
determines the density of muons.
At a density where the condition
\begin{equation}
\omega_K = \mu_n - \mu_p
\label{eq:kaonequil}
\end{equation}
is satisfied, kaons are condensed and
the kaon density $\rho_K$ becomes non-zero.
%Finally, charge neutrality is expressed as 
%\begin{eqnarray}
%\sum_B q_B \rho_B - \rho_K - \rho_e - \rho_\mu = 0,
%\label{eq:baryonconserv}
%\end{eqnarray}
%where $q_B$ is the charge of baryon $B$ and $\rho_B$ is the number
%density of $B$.
Solving the Eqs.~(\ref{sigma}--\ref{eq:kaonequil})
self-consistently and simultaneously,
%together with the quark eigenvalue equation of Eq.~(\ref{eq:bessel}), 
one can determine the 16 variables uniquely.

\section{Results}

\begin{figure}[tbp]
\begin{center}
\epsfig{file=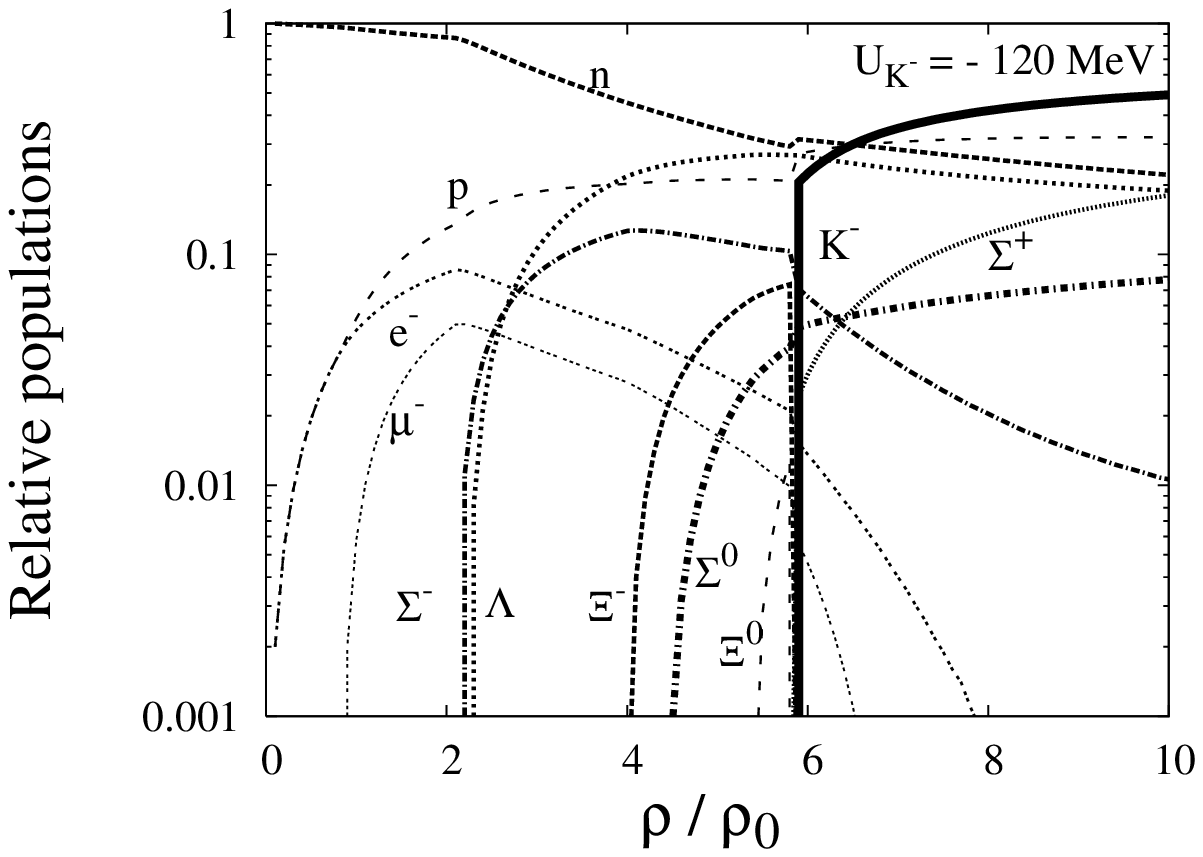, width=6.5cm}
\epsfig{file=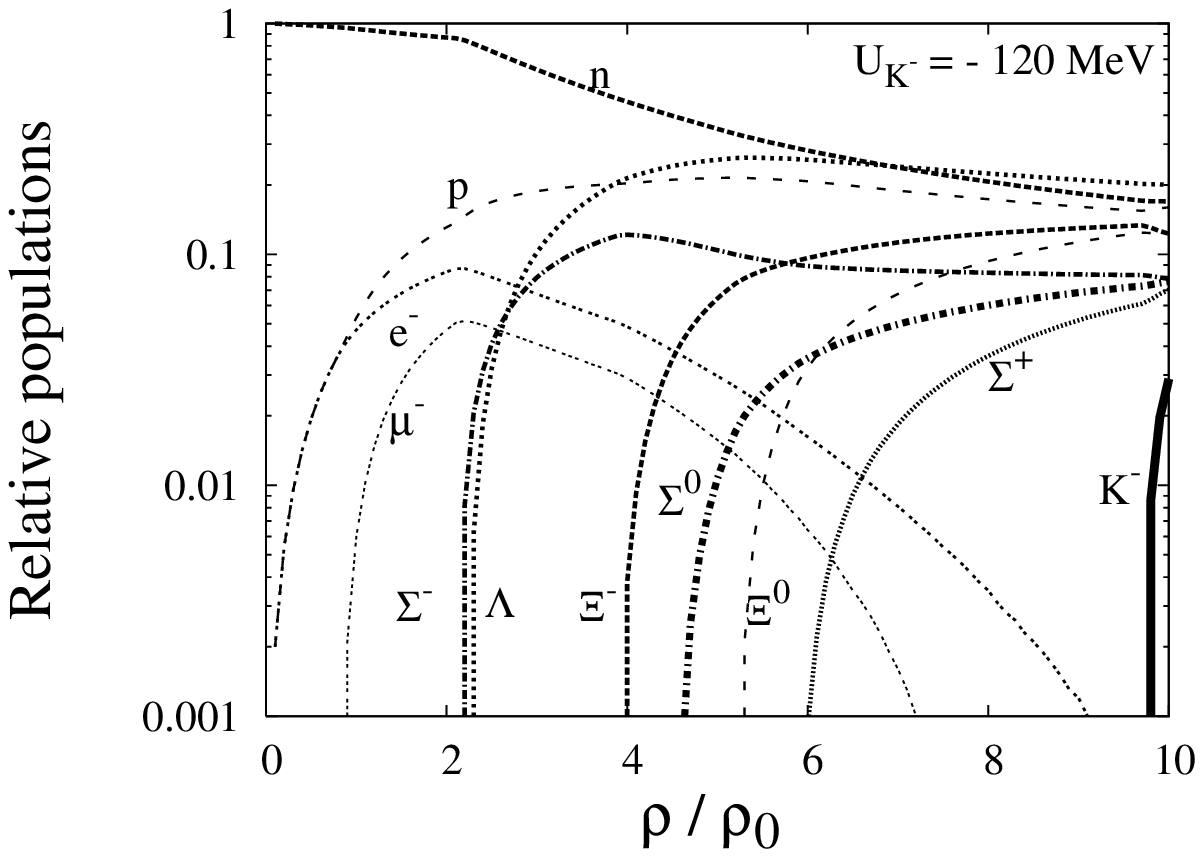, width=6.5cm} \\
\epsfig{file=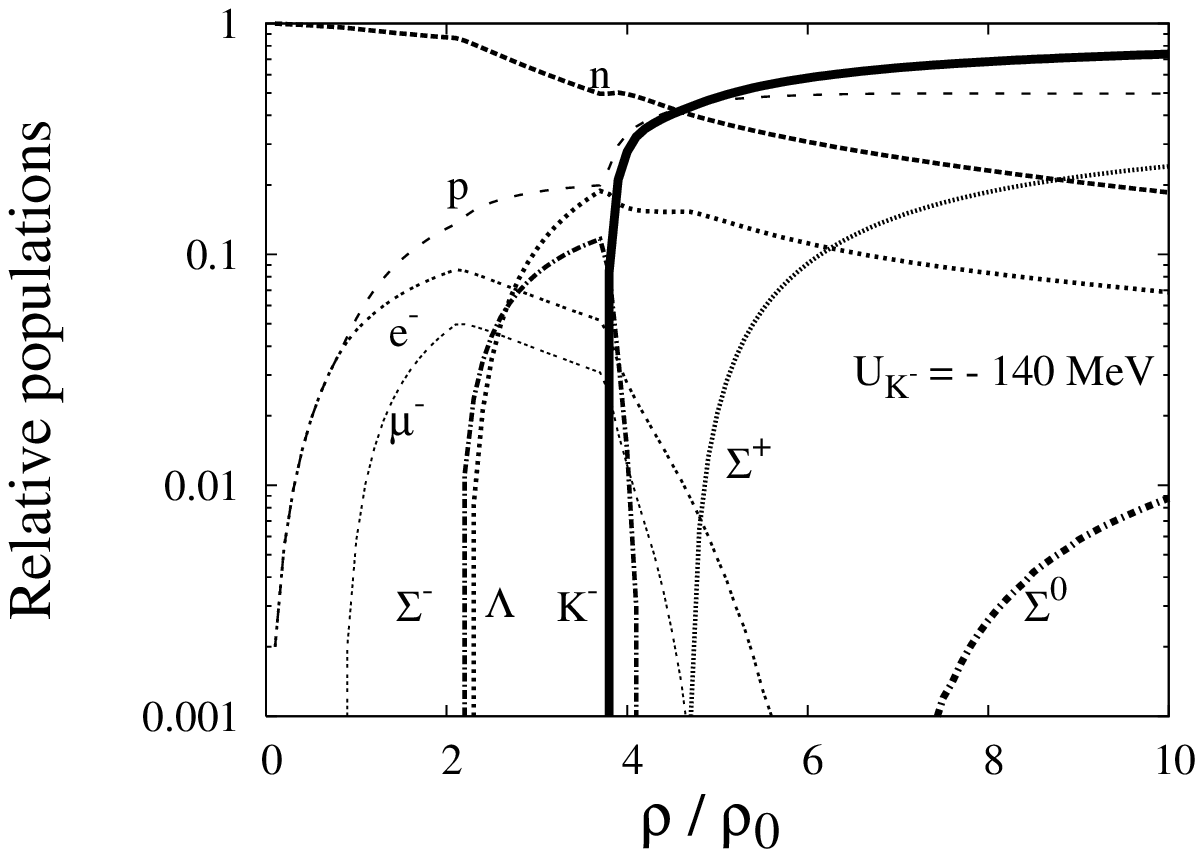, width=6.5cm}
\epsfig{file=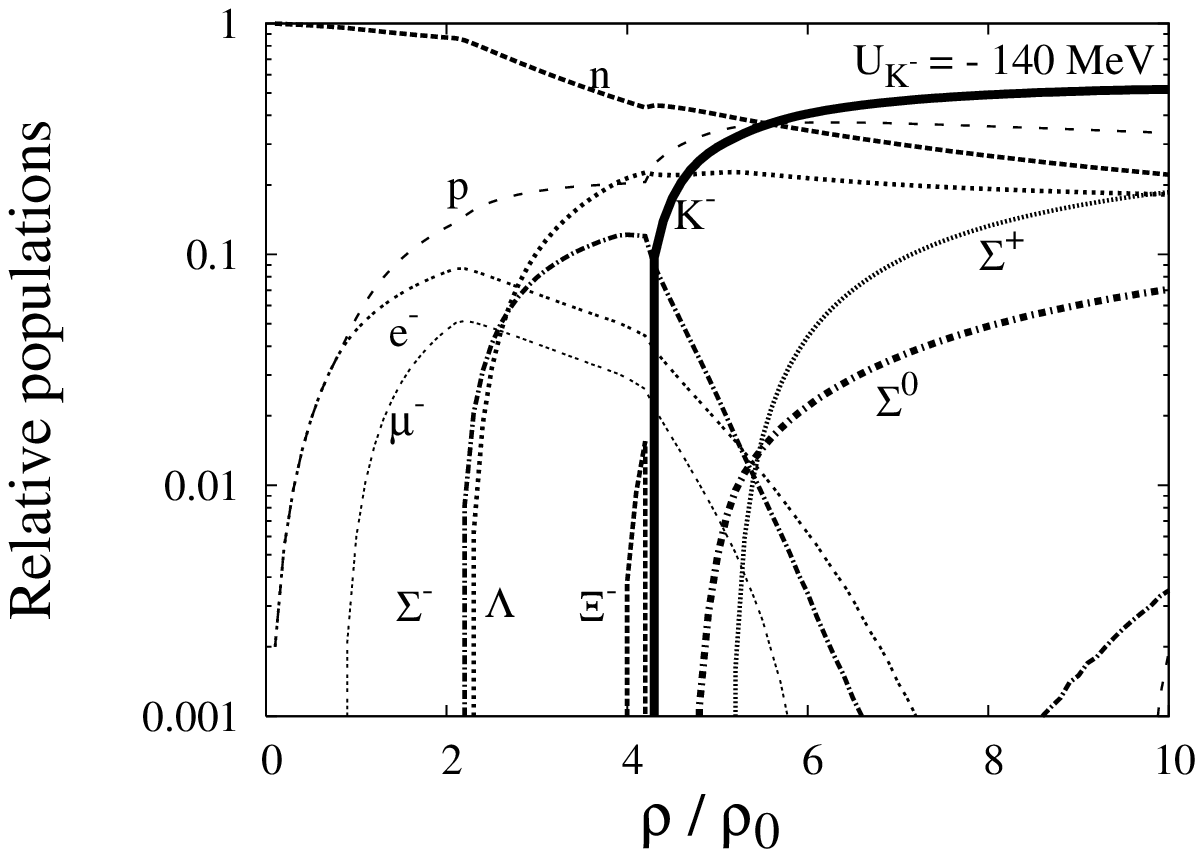, width=6.5cm}\\
\epsfig{file=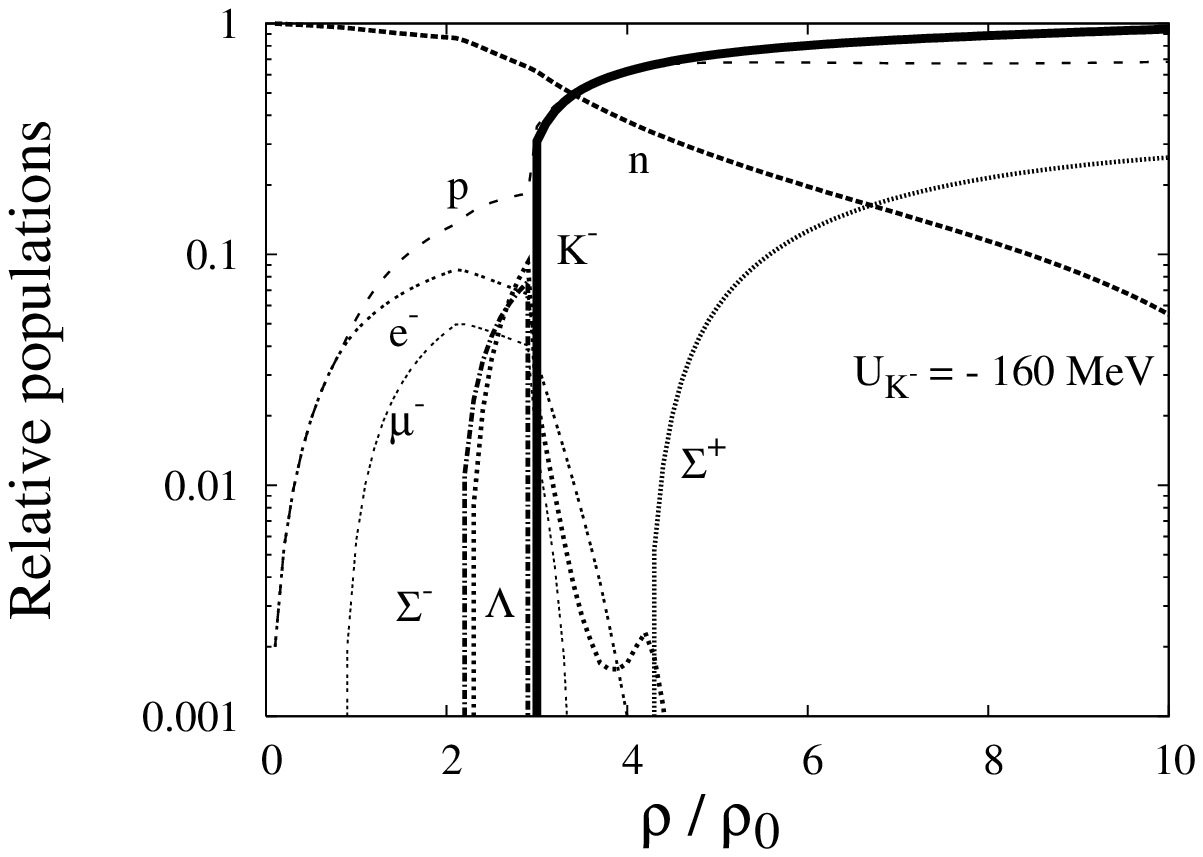, width=6.5cm}
\epsfig{file=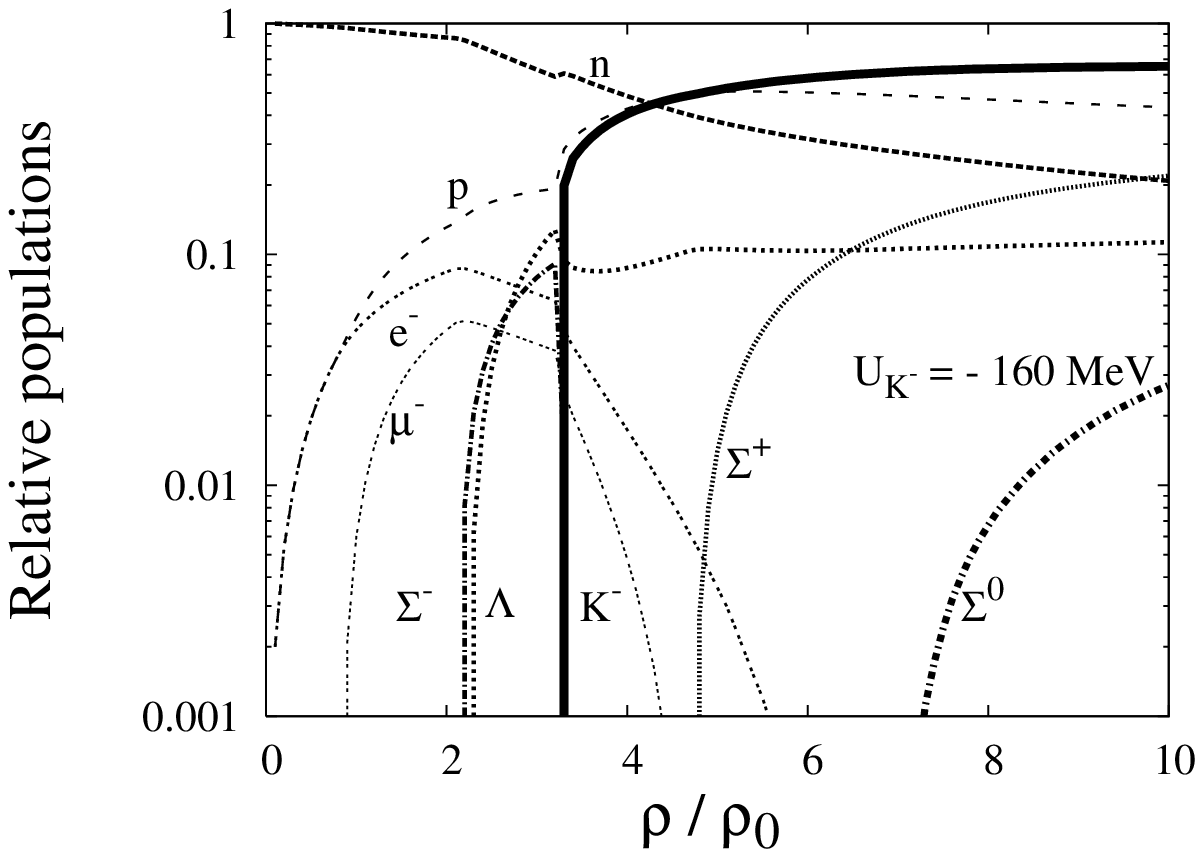, width=6.5cm}
\end{center}
\caption{Compositions of neutron star matter calculated from the
MQMC (left panels) and the QHD (right panels) models.}
\label{fig:population}
\end{figure}
Fig.~\ref{fig:population} shows the relative populations,
the ratios of the densities of octet baryons, leptons and $K^-$ to the total
baryon density, in the neutron star matter as functions of
$\rho / \rho_0$ up to $\rho=10 \rho_0$.
The left panels show the results from the MQMC model and the right 
from the QHD model for $U_{K^-} = -120, -140,$ and $-160$ MeV.
(Figures for $U_{K^-} = -80$ and $-100$ MeV are not shown here since
they are not too much different from the one for $U_{K^-} = -120$ MeV
particularly for QHD.)
Figures from both models show that the onset density of the
kaon condensation $\rho_{\rm crit}$ becomes lower as
$|U_{K^-}|$ increases.

\begin{figure}
\begin{center}
\epsfig{file=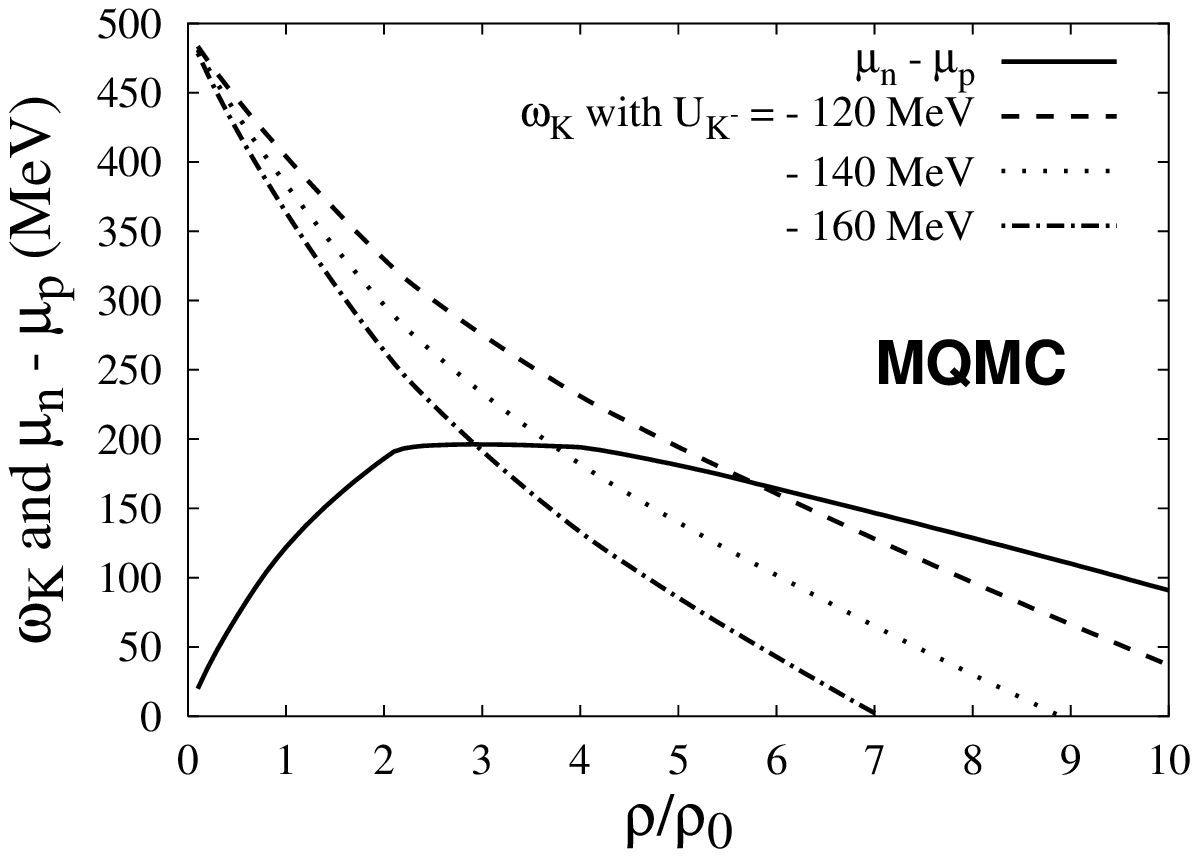, width=7cm}
\epsfig{file=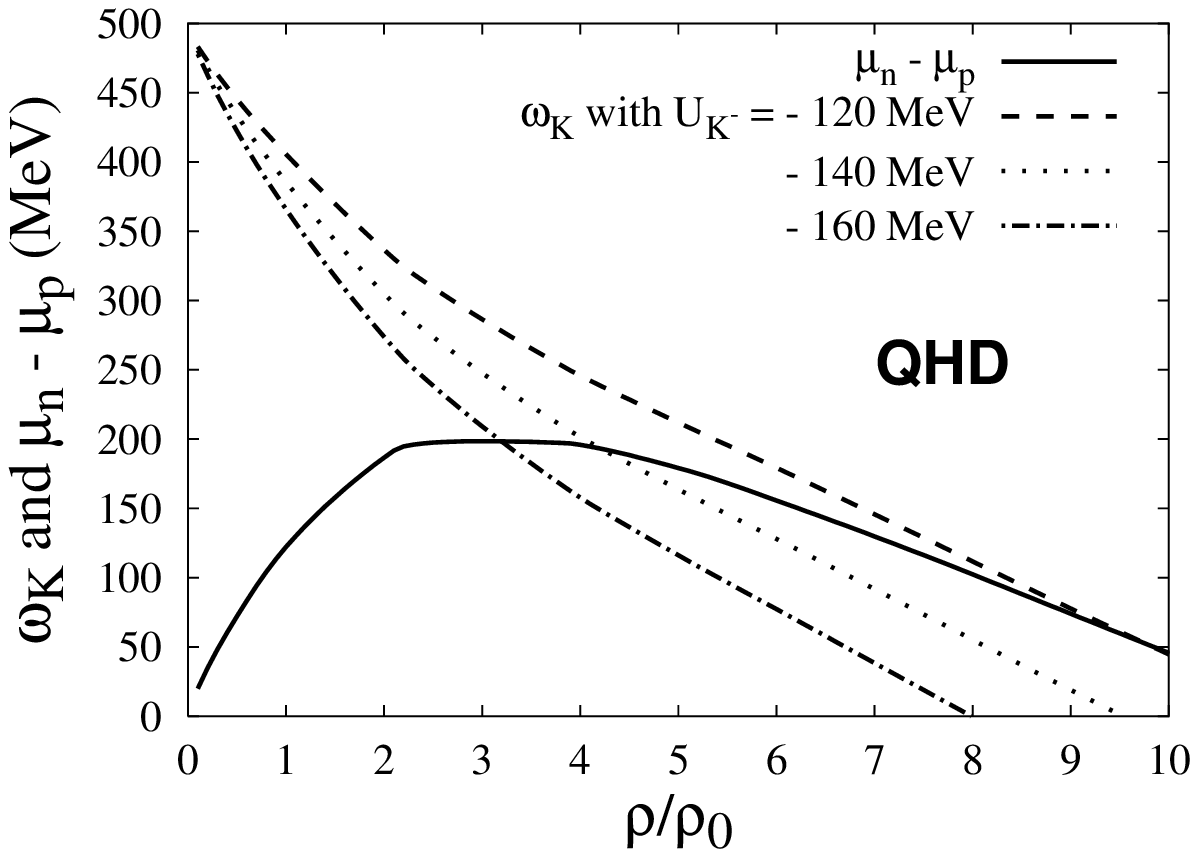, width=7cm}
\end{center}
\caption{Kaon energies $\omega_K$ for $U_{K^-}$ = $-120$ MeV (dashed),
$-140$ MeV (dotted) and $-160$ MeV (dot-dashed) 
are compared with $\mu_n - \mu_p$ (solid).
At the densities where $\omega_K$ and $\mu_n - \mu_p$ intersect, 
kaons start to condense.
The left panel is for the MQMC model, and the right for QHD.} 
\label{fig:kaonenergy}
\end{figure}

To see how $\rho_{\rm crit}$ changes depending on $U_{K^-}$,
let us consider Eq.~(\ref{eq:kaonequil}), which determines $\rho_{\rm crit}$.
%$\rho_{\rm crit}$ is determined by Eq.~(\ref{eq:kaonequil}).
Fig.~\ref{fig:kaonenergy} displays $\omega_K$ and 
$\mu_n - \mu_p$, which are, respectively,
the left and the right
hand sides of Eq.~(\ref{eq:kaonequil}) 
(computed without including kaon condensation just for producing this figure).
The left panel is from the MQMC model, and the right panel from QHD.
At a density where the curve of $\omega_K$ intersects with
that of $\mu_n - \mu_p$, kaon condensation sets in.
Among the coupling constants and meson fields that determine 
$\omega_K$, only $g_{\sigma K}$ depends on $U_{K^-}$.  
The $\sigma$-meson contributes to $\omega_K$ 
attractively, as can be seen in Eq.~(\ref{eq:kaon-dispersion}).
Thus $\omega_K$ becomes smaller 
for a larger $|U_{K^-}|$, as shown in Fig.~\ref{fig:kaonenergy}.
%Fig.~\ref{fig:kaonenergy} also shows that
%$\omega_K$ calculated from the MQMC model decreases more
%rapidly with density than $\omega_K$ from QHD 
%for each $U_{K^-}$ value.
%The curves for $\mu_n - \mu_p$ are more or less the same for both models
%at $\rho \lesssim 4\rho_0$, but at $\rho > 4\rho_0$ 
%$\mu_n - \mu_p$ decreases faster in QHD.
%Thus the intersection and kaon condensation occur at lower densities
%in the MQMC model. 
%This behaviour of the intersection in Fig.~\ref{fig:kaonenergy} 
%is well reflected in the kaon condensation onset density 
%$\rho_{\rm crit}$ in Fig.~\ref{fig:population}.

Figure~\ref{fig:population} also shows 
that as $|U_{K^-}|$ increases from $120$ MeV to $140$ MeV
$\rho_{\rm crit}$ changes drastically in both MQMC and QHD models, 
but as $|U_{K^-}|$ increases further above $140$ MeV,
$\rho_{\rm crit}$ changes only moderately.
This can also be seen from Fig.~\ref{fig:kaonenergy}.
As $|U_{K^-}|$ increases from $120$ MeV to $140$ MeV
the intersection between the curves for $\mu_n - \mu_p$ 
and $\omega_K$ moves rapidly (particularly for QHD), 
whereas when $|U_{K^-}|$ increases further above $140$ MeV
the intersection 
%between the curves for $\mu_n - \mu_p$ and $\omega_K$ 
shifts only a little to lower densities.
%%%%%
\begin{figure}
\begin{center}
\epsfig{file=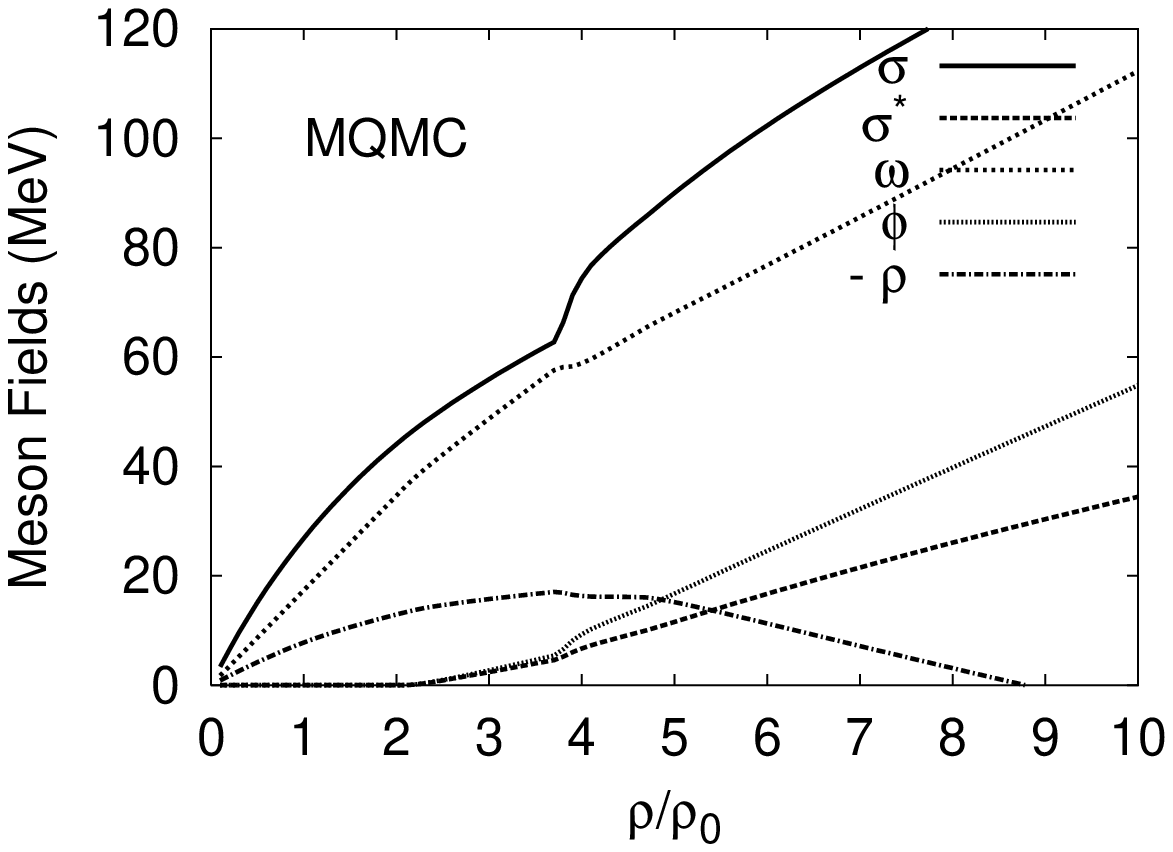, width=7.2cm}
\epsfig{file=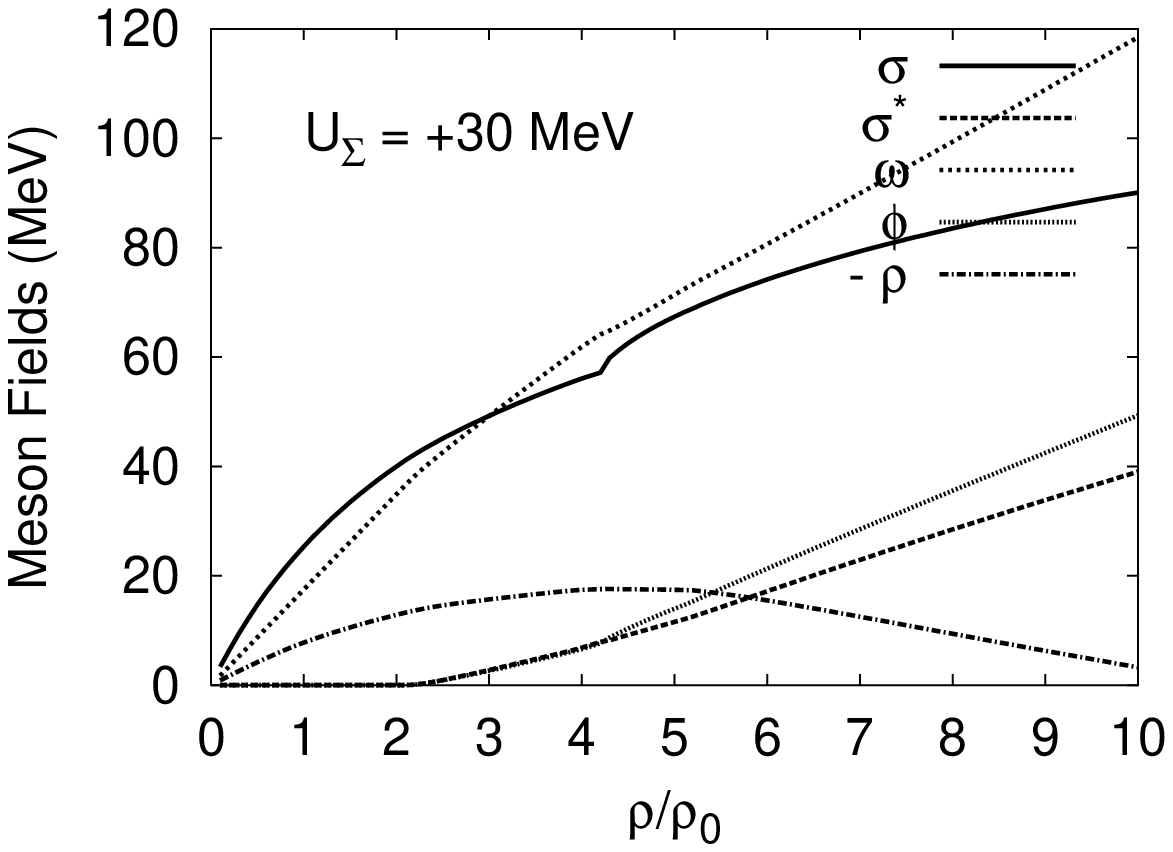, width=7.2cm}
\end{center}
\caption{Meson fields calculated from the MQMC (left) and the QHD (right)
models as functions of the matter density for $U_{K^-} = -140$ MeV.}
\label{fig:meson}
\end{figure}

Another common feature of the two models is that
regardless of $\rho_{\rm crit}$, once the kaon is created,
the density of $K^-$ piles up very quickly and overwhelms
the population of the hyperons and even the nucleons.
This behavior was also obtained by other authors
\cite{glend,kpe-prc95,bb-prc01,mpp-prc05}.
The reason can be partly attributed to the $\omega$-meson. 
The $\omega$-meson term in the energy of $K^-$ (in Eq.~(14)) 
has a negative sign and is thus attractive,
but it is repulsive for octet baryons.
Fig.~\ref{fig:meson} shows the $\omega$-meson
is a dominant meson at higher densities
in both MQMC and QHD models. 
Thus the $\omega$-meson enhances the population of $K^-$ 
but suppresses baryons, and the kaon density increases rapidly.
%In Fig.~\ref{fig:meson}, we plot the meson fields calculated
%from the MQMC and QHD models as functions of 
%matter density $\rho$.
%Since the $\omega$-meson contribution to $\omega_K$ is 
%smaller than the hyperons by the order of $g^q_\omega \omega_0$, 
%once the kaon is condensed, the ground state strongly
%favors kaons rather than the hyperons, and 
%the kaon density increases rapidly.
In addition, due to the competition between
the negatively charged hyperons and $K^-$
in the charge neutrality condition, the negative hyperons are highly 
suppressed and in some cases not even created at all 
as soon as the kaon condensation sets in.
Positively charged hyperons, on the other hand, receive the 
opposite effects from the kaon condensation, and  
$\Sigma^+$ is created at lower densities as $|U_{K^-}|$ increases more.
The proton density is also enhanced by large abundance of $K^-$,   
which facilitates in turn the enhancement of $\Sigma^+$ population
through the chemical equilibrium condition
of the positively charged hyperons in Eq.~(\ref{eq:chemi-eq}).
\begin{figure}
\begin{center}
\epsfig{file=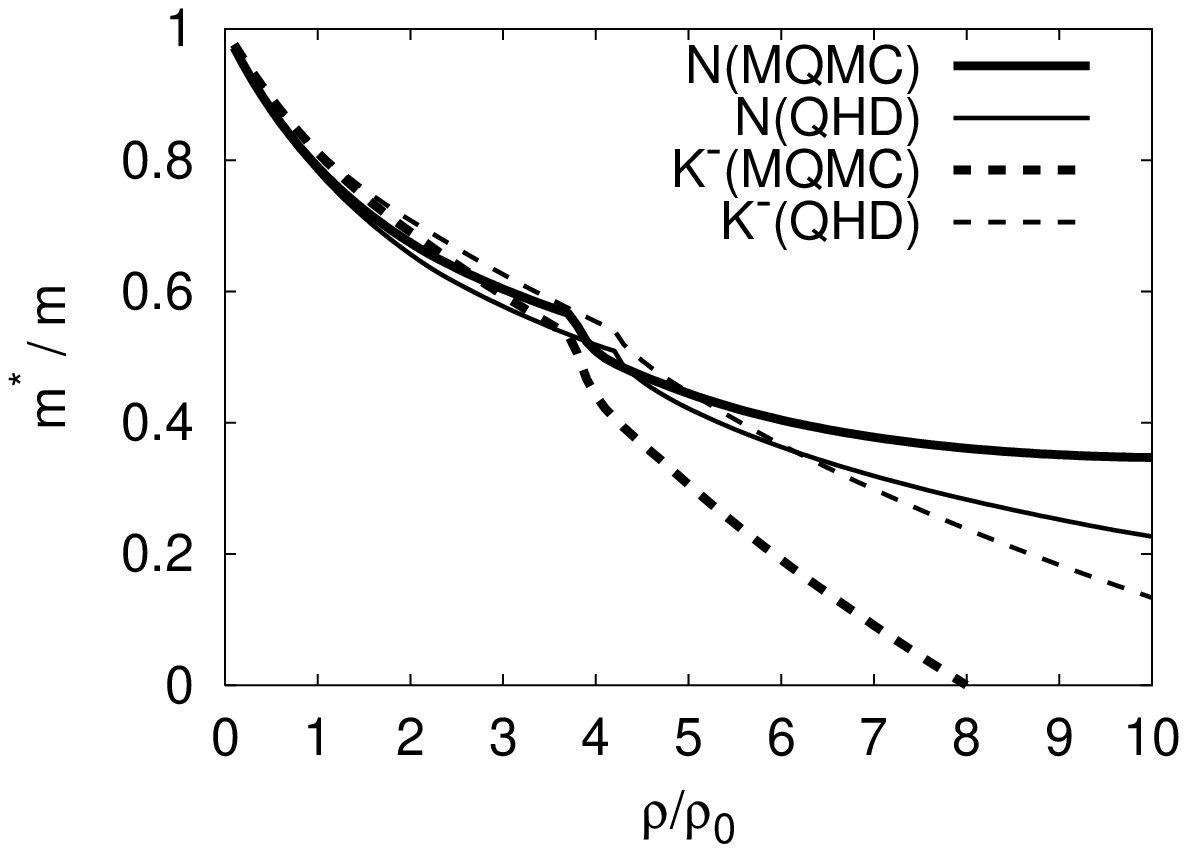, width=7.2cm}
\epsfig{file=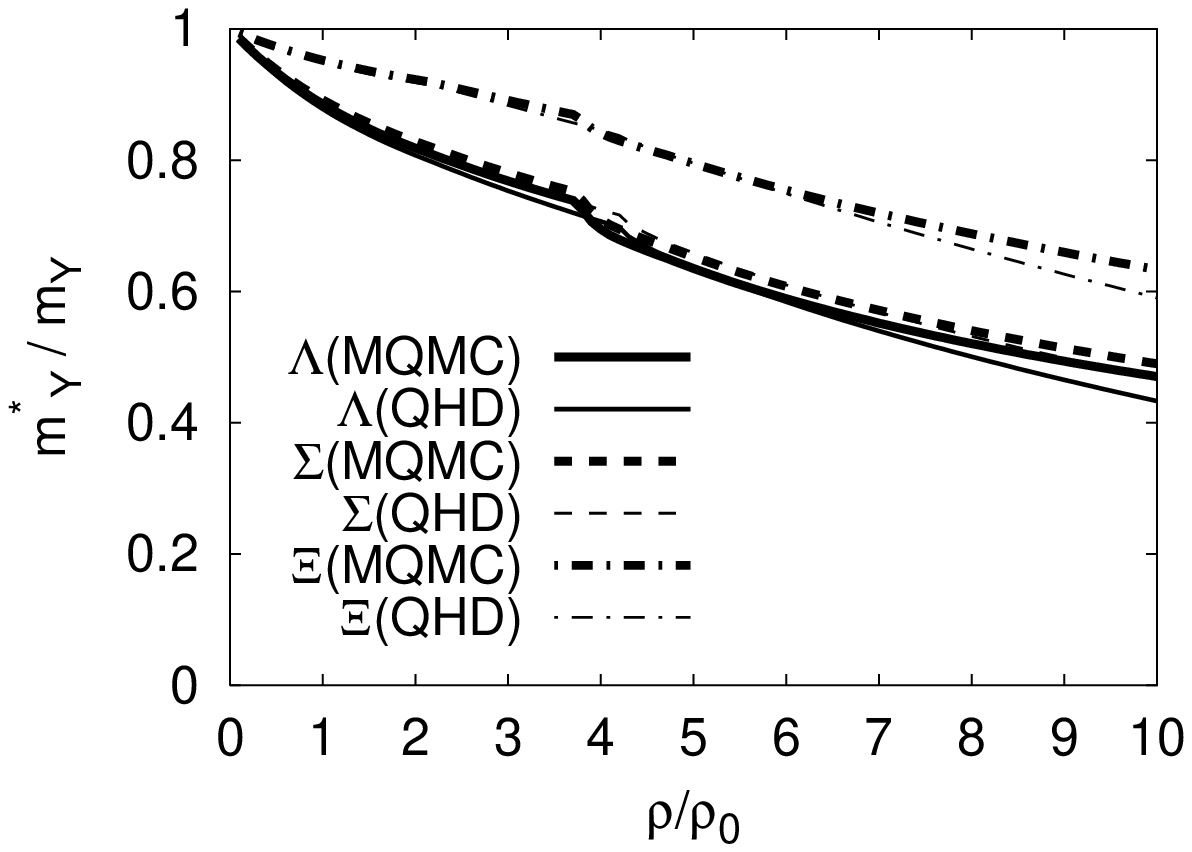, width=7.2cm}
\end{center}
\caption{Effective masses of the nucleons, kaon, 
and hyperons are plotted for $U_{K^-} = - 140$ MeV.}
\label{fig:mNstar}
\end{figure}

Let us now discuss different aspects of the two model calculations.
First, Fig.~\ref{fig:population} shows that $\rho_{\rm crit}$ from 
the MQMC model is lower than that from QHD.
For $U_{K^-} = -120, -140$, and $-160$ MeV, $\rho_{\rm crit}$
values are $5.9 \rho_0$, $3.8 \rho_0$ and $3.0 \rho_0$ 
in the MQMC model, respectively, while they are
$9.8 \rho_0$, $4.3 \rho_0$ and $3.3 \rho_0$ in QHD.
Secondly, the MQMC model predicts a larger population of the kaon
than QHD for a given $U_{K^-}$ value.
%$\rho_{\rm crit}$ is determined 
%by Eq.~(\ref{eq:kaonequil}).
%As mentioned earlier in explaining  
%Fig.~\ref{fig:kaonenergy},
%$\omega_K$ from the MQMC model is smaller than $\omega_K$ from QHD.
Figure~\ref{fig:kaonenergy} shows that
$\omega_K$ calculated from the MQMC model decreases more
rapidly with density than $\omega_K$ from QHD 
for each $U_{K^-}$ value.
The curves for $\mu_n - \mu_p$ are more or less the same for both models
at $\rho \lesssim 4\rho_0$, but at $\rho > 4\rho_0$ 
$\mu_n - \mu_p$ decreases faster in QHD.
Thus the intersection and kaon condensation occur at lower densities
in the MQMC model. 
This behaviour of the intersection in Fig.~\ref{fig:kaonenergy} 
is well reflected in the kaon condensation onset density 
$\rho_{\rm crit}$ in Fig.~\ref{fig:population}.
Fig.~\ref{fig:meson} shows that the 
$\sigma$-meson field calculated by the MQMC model is larger than that
calculated by QHD.
A larger $\sigma$-field in the MQMC model makes $m^*_K$ and consequently 
$\omega_K$ smaller.
On the other hand, as seen in Fig.~\ref{fig:kaonenergy},
$\mu_n - \mu_p$ from QHD decreases faster with density
at higher densities than that from MQMC.
Thus the intersection of the $\omega_K$ curve with the curve for $\mu_n - \mu_p$
occurs at lower densities with the MQMC model.
%, as shown in Fig.~\ref{fig:kaonenergy}. 
Therefore, $\rho_{\rm crit}$ is smaller in the MQMC model.

Another model dependency of the results can be seen from   
the population of kaons, which is larger in the MQMC model.
The effective mass of a kaon as a point particle is 
determined by $\sigma$ and $\sigma^*$ mesons through the relation 
$m_K^* = m_K - g_{\sigma K}\sigma -g_{\sigma^* K} \sigma^*$
and is plotted in Fig.~\ref{fig:mNstar}. 
Since the $\sigma$ fields are larger in the MQMC model 
(as shown in Fig.~\ref{fig:meson}), the effective mass and 
the energy of a kaon are smaller in the MQMC model than QHD.
Thus, kaon condensation takes place more in the MQMC model.

Figures~\ref{fig:meson} and \ref{fig:mNstar} also show that
even though $\sigma$-meson field from the MQMC model is larger  
than that from QHD as the densities increase, 
the reduction of the effective mass of baryons  
is smaller (or similar) with the MQMC model.
If one could parametrize the effective mass of baryons from 
the MQMC model in the form of
$m_B^* = m_B - g_{\sigma B}(\sigma) \sigma 
-g_{\sigma^* B}(\sigma^*) \sigma^*$
where $g_{\sigma B}(\sigma)$ and $g_{\sigma^* B}(\sigma^*)$ are 
functions of $\sigma$ and $\sigma^*$, respectively,
the results in Fig.~\ref{fig:mNstar}
might imply that $g_{\sigma B}(\sigma)$ and 
$g_{\sigma^* B}(\sigma^*)$ are decreasing functions 
with respect to the density.
The rate of decrease is rather high since the
product $(g_{\sigma B}(\sigma)\, \sigma)_{\rm MQMC}$
is smaller than (or similar to) $(g_{\sigma B}\, \sigma)_{\rm QHD}$ 
while the $\sigma$-field value from MQMC is much greater than 
that from QHD.
Such a decrease of $g_{\sigma B}(\sigma)$ in the MQMC model
may be regarded as partial restoration of the chiral symmetry
at high densities.
%Thus the larger values of $\sigma$ field in MQMC is created
%to keep the equivalence between the attractive ($g_{\sigma B}(\sigma)\sigma$)
%and the repulsive potential ($g_{\omega B}\omega$) 
%from the reduced $g_{\sigma B} (\sigma)$ 
%and make kaon condensation stronger.

We have calibrated both the MQMC and QHD model parameters to the same
saturation properties. However, we find that the neutron star matter composition
profiles from the two models are quite different and
that they show significant model dependence.
QHD assumes the baryons as point particles, 
whereas the MQMC model treats the baryons as MIT bags. 
Thus, the major difference between the two models is in the 
definition of the  
effective mass of baryons, $m^*_B$. 
The equation of motion for the $\sigma$-meson field
is also different accordingly.
$m^*_B$ in QHD is a simple linear function of 
the $\sigma$-field, and the factor $C_B(\sigma)$ 
in Eq~(\ref{sigma}), is a constant.
In the MQMC model, $m^*_B$ is a
non-linear function of $\sigma$-field, and thus 
$C_B(\sigma)$ is highly non-linear.
When these non-linear $m^*_B$ and $C_B(\sigma)$ are expanded
in powers of the $\sigma$-field, an infinite number of $\sigma$-field
terms would appear. 
(Cubic and quartic terms are explicitly taken into account
in the QHD model as in Eq.~(\ref{eq:U_QHD}).)
Higher order terms can be interpreted as higher order contributions
such as self interactions of meson fields,
which are believed to be more important at high densities.
But at high densities it can be questioned 
whether the non-linear terms of the
$\sigma$-meson in the MQMC model 
account for the higher order effects properly and consistently.
For instance, it is generally known that as the baryons come closer
to each other the interplay of heavy mesons becomes more important.
At high enough densities, their self interaction contributions may need
to be included on the same ground as for the $\sigma$-meson,
but the present MQMC model truncates the heavy
meson terms at the leading order. 

It seems worthwhile to discuss at this point two more aspects 
of our results. The first
one is the EoS and the resulting mass-radius relation of the
neutron star. The second point is
the dependence of our results on the $\Sigma$ hyperon interaction
in matter, which is not well known yet. 

\begin{figure}[tbp]
\begin{center}
\epsfig{file=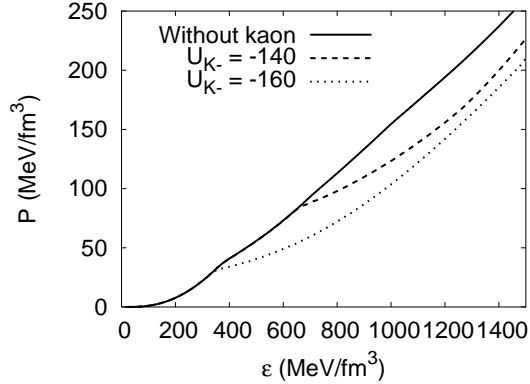, width=7.5cm}
\end{center}
\caption{Comparison of the EoS with and without kaons in QHD model.
The Gibbs condition is used to treat the mixed phase.}
\label{fig:eos_comp}
\end{figure}

Let us first consider the EoS and the maximum mass of the neutron star.
As the kaon ($K^-$) appears and condensates, 
the number of negatively-charged hyperons decreases to satisfy the
charge neutrality.
The decrease in the number of baryons will result in the reduction
of the pressure and lead to softnening the EoS. 
Fig.~\ref{fig:eos_comp} shows such a softening of the EoS due to the 
kaon condensation. 
In calculating the EoS of a system consisting of multicomponent 
substances, as in the case with the kaon condensation, 
Gibbs condition has to be employed for the proper description of the 
mixed phase.
The curves in Fig.~\ref{fig:eos_comp}
are the results obtained with the QHD and Gibbs condition. 
As kaons appear, the EoS becomes considerably soft and the effect
becomes more pronounced with a stronger attraction, 
i.e., for a larger $|U_{K^-}|$ value.
For the MQMC model, however, applying the Gibbs conditions do not
give us a converging solution. Solving the 16 highly nonlinear 
equations together with Gibbs conditions double the number of equations
to be solved, and the convergence could not be reached.
It is not clear to us whether the convergence problem is due to
numerical problems or due to non-linearity which can cause 
bifurcation or chaos. 
Therefore, we have used Maxwell construction for the MQMC model.
(Some literatures \cite{Maxwell} show that Maxwell construction
is a good approximation to the Gibbs condition, but in some other 
literature \cite{glend} it was emphasized that Gibbs condition 
produces significantly different results from those of Maxwell 
construction. Below we show that in our case the neutron star mass
itself does not change much whether we use Maxwell or Gibbs conditions
for QHD. Thus our use of Maxwell construction for the MQMC model
may be considered as an acceptable approximation.)
We solve Tolman-Oppenheimer-Volkoff equation 
to calculate the maximum mass of a neutron star. 
The results are shown in Tab.~\ref{tab:mass-radius},
where the central density, the maximum mass, and the corresponding radius
are listed for both MQMC and QHD models.
%
%The MQMC model, using the Maxwell construction gives us 
%the maximum of neutron star to be
%1.611 $M_\odot$, 1.526 $M_\odot$ and 1.453 $M_\odot$ 
%for $|U_{K^-}| = $120, 140, 160 MeV, respectively.
%The QHD model with Maxwell construction predicts the maximum mass
%to be 1.503 $M_\odot$, 1.460 $M_\odot$ and 1.322 $M_\odot$
%for $|U_{K^-}| = $ 120, 140, 160 MeV, respectively.
%If we use the Gibbs condition for the QHD model, 
%the maximum mass of neutron star becomes
%1.503 $M_\odot$, 1.452 $M_\odot$ and 1.190 $M_\odot$ 
%for $|U_{K^-}| = $120, 140, 160 MeV, respectively,
%which are essentially the same as the masses obtained with
%the Maxwell construction.
%
\begin{table}
\begin{center}
\begin{tabular}{c|ccc|ccc|ccc} \hline
 & \multicolumn{3}{c|}{MQMC (Mx)} & \multicolumn{3}{c|}{QHD (Mx)} &
\multicolumn{3}{c}{QHD (Gb)} \\
\hline
$U_{K^-}$ &
$\rho_c/\rho_0$ & $M/M_\odot$ & $R$ &
$\rho_c/\rho_0$ & $M/M_\odot$ & $R$ &
$\rho_c/\rho_0$ & $M/M_\odot$ & $R$ \\
\hline
$-120$ & 6.2 & 1.61 & 11.8 & 6.1 & 1.50 & 11.4
& 6.1 & 1.50 & 11.4 \\ \hline
$-140$ & 4.6 & 1.53 & 12.8 & 5.0 & 1.46 & 12.1
& 5.0 & 1.45 & 12.1 \\ \hline
$-160$ & 4.6 & 1.45 & 13.1 & 4.0 & 1.32 & 12.7
& 4.3 & 1.19 & 12.3 \\
\hline
\end{tabular}
\end{center}
\caption{The maximum mass of a neutron star $M$, the corresponding radius $R$
and the density at the center of the star $\rho_c$ are listed for
$U_{K^-} = -120, -140,$ and $-160$ MeV. $U_{K^-}$ is in units of MeV,
and $R$ in km. ``Mx" and ``Gb" refer to Maxwell and Gibbs
conditions, respectively.}
\label{tab:mass-radius}
\end{table}
The maximum mass calculated with QHD model is roughly 10 \% smaller than
that with MQMC model. For both models the maximum mass becomes smaller
for a larger $|U_K |$. 
The maximum mass calculated with MQMC and $|U_{K^-}| = $ 160 MeV
is compatible with observation, while
the maximum mass calculated with QHD and $|U_{K^-}| = $ 160 MeV 
becomes too small to be compatible with the observed values.
However, this fact may not necessarily rule out the possibility of
$|U_{K^-}| $ becoming as large as 160 MeV 
because there are other possible mechanisms
which are not included.

\begin{figure}[tbp]
\begin{center}
\epsfig{file=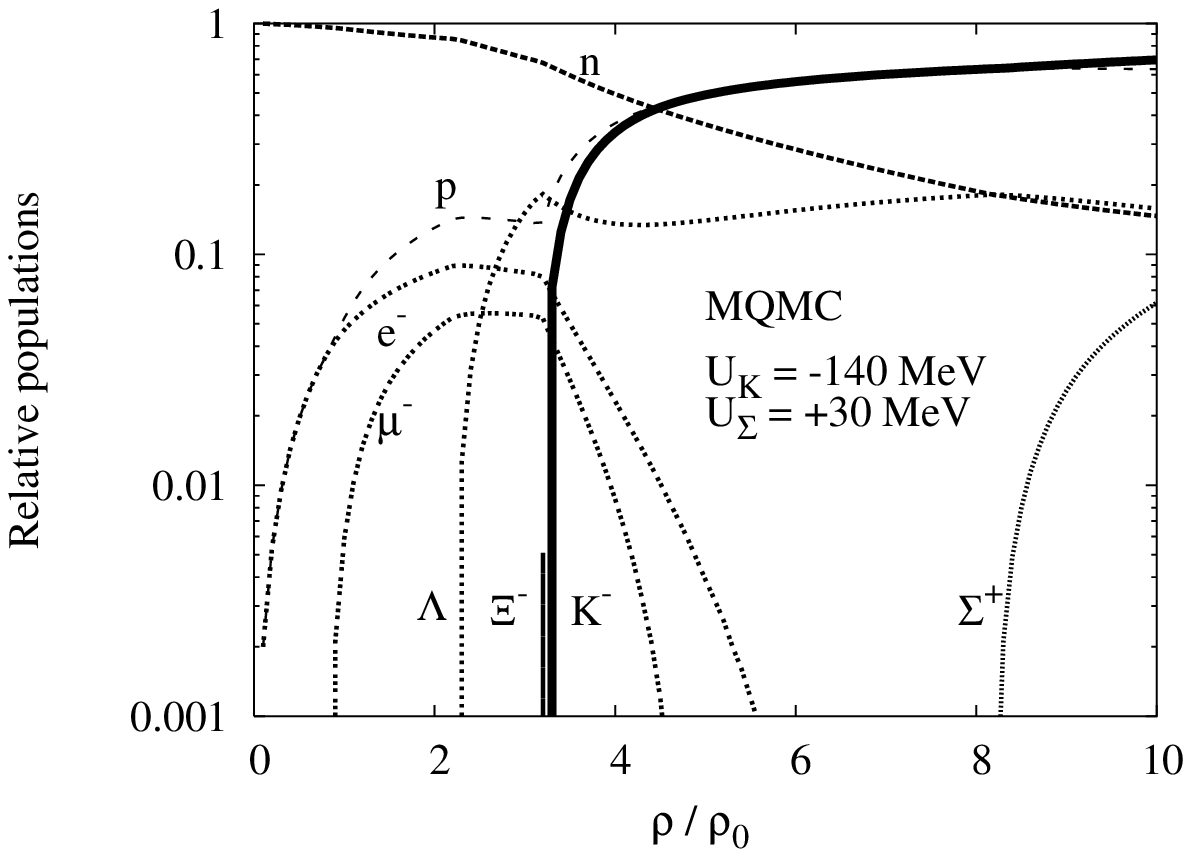, width=6.5cm}
\epsfig{file=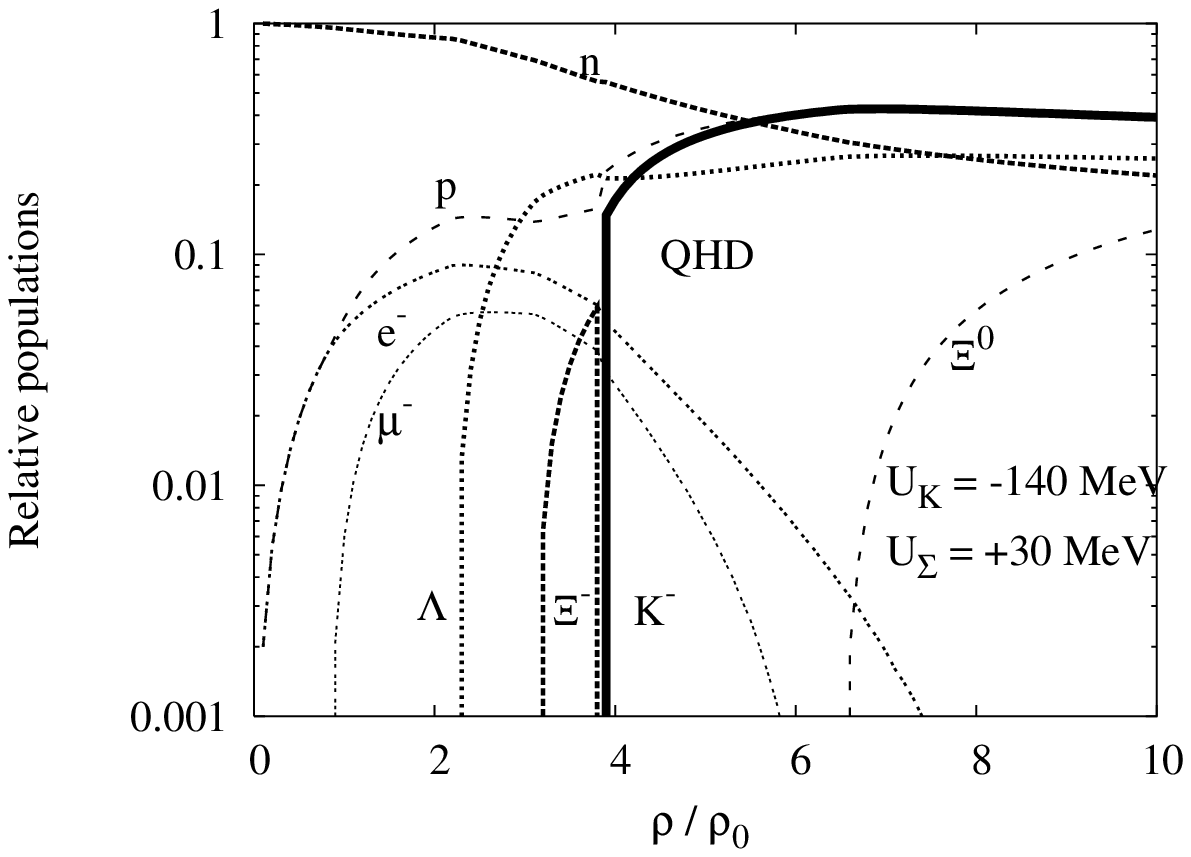, width=6.5cm}
\end{center}
\caption{Comparison of the population from the MQMC (left) with
that of the QHD (right). 
The calculations are done with $U_\Sigma = +30$ MeV and $U_{K^-} = -140$ MeV.}
\label{fig:ppl_repul_Sigma}
\end{figure}

Now let us consider the second aspect mentioned above.
In the calculations made so far, we have assumed quark-counting rule 
in determining the hyperon-meson coupling constants. 
Experiments on $\Lambda$-hypernuclei indicate that quark counting is a good
approximation of the realistic interaction of $\Lambda$ hyperons in nuclei,
which gives the value of $\Lambda$ optical potential at saturation 
density in the range $-40 \sim -30$ MeV.
On the other hand, there is large ambiguity in the $\Sigma$ hyperon 
interaction strength. Ref.~\cite{npa95} shows that $\Sigma$ hyperon 
feels repulsion rather than attraction in nuclear medium. 
%and suggests optical potential value is in the range $+10 \sim +30$ MeV.
There are also experimental indications that the $\Sigma$ hyperon
interaction is repulsive \cite{saha}.
%However, the interaction of the $\Sigma$ hyperon in dense medium
%requires further investigation.
If the $\Sigma$ interaction is indeed repulsive, the population profile
of neutron star matter can change
significantly from what is shown earlier in this work since
we have used the quark counting rule.
The number of $\Sigma^-$ is closely related to the onset of
$K^-$ condensation since they compete with each other for
the charge neutrality condition.
To see the effect of possible repulsive nature of $\Sigma$ interaction
on the kaon condensation, we have repeated the calculations
with repulsive $\Sigma$ interaction. 
We first fit the coupling constants $g'^{\Sigma}_\sigma$ in MQMC and 
$g_{\sigma \Sigma}$ in QHD so that the $\Sigma$ optical potential value
at the saturation density equals to $+30$ MeV, and fix the remaining 
meson-$\Sigma$ coupling constants with the quark counting rule.
The resulting population profiles with the kaon optical potential 
$U_{K^-} = -140$ MeV are shown in Fig.~\ref{fig:ppl_repul_Sigma}.
Compared to the quark-counting results in Fig.~\ref{fig:population},
the onset of kaon condensation occurs at slightly lower densities. 
This minor change happens regardless of $U_{K^-}$ value. 
However, the main features of
kaon condensation, i.e., its onset density, fast increase of population
and its dominance at high densities are not much affected by the
change of $\Sigma$ interaction in nuclear medium. 

\section{Conclusions and discussions}

Using the modified quark-meson coupling model,
we have obtained the composition profiles of neutron star matter, 
focusing on the effects of the strange particles of hyperons and kaons.
Motivated by recent theoretical predictions of deeply bound kaonic states
\cite{akaishi} and the subsequent claims of the observations of 
interesting peaks found 
in KEK\cite{s0, suzuki1}, FINUDA\cite{FINUDA}, and BNL\cite{kishi-npa05}  
experiments, large kaon optical potential $U_{K^-}$ was considered.
By varying the value of $U_{K^-}$, we have investigated how the 
onset density of the kaon condensation and
the composition of the stellar matter change.
Employing the QHD model parameters which satisfy exactly 
the same saturation conditions
as the MQMC model, we have investigated 
the model dependence of the results.

We observed two common features from the two model calculations.
First, a larger $|U_{K^-}|$ produces a smaller onset density
of the kaon condensation.
This behavior is easily understood from the relation between
$U_{K^-}$ and $g_{\sigma K}$ together with the role of $g_{\sigma K}$
to the energy of the kaon $\omega_K$.
Secondly, the number of kaons rapidly increases, 
and the number of negatively charged hyperons is strongly suppressed.
This is due to the fact that the $\omega$-meson 
gives rise to attraction to $K^-$
whereas it couples to baryons repulsively.

Model dependence was also observed.
The kaon condensation takes place at lower densities
in the MQMC model. 
The number of kaons is always larger with the MQMC model 
for given $U_{K^-}$ values.
Larger $\sigma$-meson fields in the MQMC model can explain 
these behaviors.
%and in the QHD model smaller value of the effective mass of hyperons.
The differences in the results from the two models become more prominent
at larger densities.
Growing discrepancies at higher densities have its
origin partly in the effective mass of baryons in
each model, which greatly affects the self-consistency
condition of the $\sigma$-meson.
The factor $C_B(\sigma)$ in the self-consistency condition
of the $\sigma$-meson is highly non-linear in the MQMC model,
which can be interpreted as 
an infinite number of $\sigma$-meson self-interaction terms. 
These higher order terms may require more proper and consistent treatment 
at high densities.
%, and more information on the nuclear matter at
%either saturation or higher densities.
%{\bf
%Chiral symmetry is an important ingredient that governs
%the strong dynamics.
%If one can incorporate the chiral symmetry in the
%description of dense matter at the level of
%quark degrees of freedom, it can provide
%a useful guide to the understanding of the role of
%higher order corrections at high densities.
%}

An important issue in the dense matter physics is the restoration 
of the chiral symmetry.
According to Ref.~\cite{br-prl91}, not only the
mass but also the pion decay constant and meson-nucleon coupling
constants decrease at a similar ratio at around the
nuclear saturation density.
In Ref.~\cite{rhhj-epja05} 
the idea of scaling behaviour is applied to the neutron star matter
using MQMC and QHD models
with only nucleon degrees of freedom, and it is shown that 
the equation of state becomes stiffer when 
scaling effects are considered.
This implies that if we include a scaling behaviour in our present models,
it may ignite the onset of exotic states earlier than the present results
which do not include a scaling.

In the kaon sector, 
the coupling constants of a kaon and exchange mesons
are currently an important issue.
We took various values of the optical potential of $K^-$ as an input
to fix $g_{\sigma K}$.
Other kaon-meson coupling constants are fixed from naive quark counting.
%The optical potential of $K^+$, 
%$U_{K^+} = - g_{\sigma K} \sigma + g_{\omega K} \omega_0$, thus
%determined turns out to be attractive at the saturation density, 
%as shown in Fig.~\ref{fig:ukplus}.
It is known, however, that $K^+$ potential is repulsive 
with the magnitude $U_{K^+} \sim 10$ MeV 
at the saturation density \cite{llb-prl97}.
If $U_{K^+}$ as well as $U_{K^-}$ is used as an input, $g_{\sigma K}$ and
$g_{\omega K}$ can be determined uniquely.
For instance, if we take $U_{K^-} = -120$ MeV and $U_{K^+} = 20$ MeV,
then we get $g_{\sigma K} = 2.041$ and $g_{\omega K} = 4.187$.
$g_{\sigma K}$ becomes smaller than the value listed in Table IV, 
while $g_{\omega K}$ becomes nearly twice of $g_{\omega K}$
fixed from the quark counting.
Both $\sigma$ and $\omega$ mesons contribute to the $K^-$ energy 
attractively, but since the $\omega$ meson becomes a dominant 
component at higher densities, taking into account of $U_{K^+}$ can produce
appreciably different results.
It may be interesting to see the effects of $U_{K^+}$
on the kaon condensation.

In our calculations, we have assumed the kaon as a point particle in both 
quark and hadron models. 
Comparison of the two models, however, shows that whether we treat 
a hadron as a bag (MQMC) or a point particle (QHD) can produce a significant 
difference.
Therefore, it is worthwhile to treat the kaon as a bag and 
compare the corresponding result with that of a point-like kaon.
In Ref.~\cite{mpp-prc05} a kaon is treated as a bag in the framework
of the QMC model, but no work has been done yet with the MQMC model.

We assume $m^*_K$ as a linear function of 
$\sigma$-field, but 
some authors employ a non-linear form \cite{kpe-prc95, j-schaf}:
\begin{equation}
m^{*2}_K = m^2_K - g_{\sigma K} m_K \sigma.
\end{equation}
If we expand this expression in powers of $\sigma/m_K$, we obtain
\[
m^*_K \simeq m_K \left[
1 - \frac{1}{2} g_{\sigma K} \frac{\sigma}{m_K} + O(\sigma^2/m^2_K)
\right].
\]
The leading order term of the $\sigma$-field has a factor 1/2,
which is not present in Eq.~(12).
Due to the factor 1/2, 
the rate of decrease in $m^*_K$ with density 
would be reduced by a factor 2,
and this would shift the kaon condensation onset density to higher densities.
This dependence on the kaon Lagrangian may be worthwhile to be studied.
%
%\begin{figure}
%\begin{center}
%\epsfig{file=ukplus.eps, width=8cm}
%\end{center}
%\caption{$U_{K^+}$ from the QHD model. At the saturation density,
%$U_{K^+}$ is attractive regardless of the $U_{K^-}$ value.}
%\label{fig:ukplus}
%\end{figure}
%

\section*{Acknowledgments}
SWH thanks B. K. Jennings and TRIUMF for hospitality during his sabbatical
leave.
This work was supported by Korea Research Foundation Grant
funded by Korea Government (MOEHRD, Basic Research Promotion Fund)
(KRF-2005-206-C00007).

\end{document}